# A STUDY ON THE SECTORS OF ECONOMY SERVICED BY PRE-INDUSTRY SYSTEM DEVELOPERS AMONG COMPANIES IN METRO MANILA: A TOOL FOR BUSINESS REENGINEERING

Engr. Daniel D. Dasig, Jr.

Professor, College of Computer Studies and Engineering, Jose Rizal University
Mandaluyong City, Philippines

## ABSTRACT

*In the emergence of transformative global economy, information system has became a necessity in businesses to obtain organizations operational excellence, adaptation to new business models, improved decision making and providing exceptional customer service, and eventual competitive advantage of the enterprise setting while keeping business alliances. This paper presents sectors of economy serviced by the pre-industry developers, explores the evolution of computer-based information system designed and developed by pre-industry system developers, and examine the effects of an information system in business to countervail indentified recurring problems. Nineteen of forty-six identified sectors of economy falls in the categories of primary, secondary, tertiary, quarternary and quinary were the recipient of computer-based system designed and developed. There have been several effects of computer-based systems to organizations, including the implied relevance to their business processes, continuum process improvement, business process reengineering, business driver and facilitator, and customer satisfaction.*

## 1. INTRODUCTION

Managing business operational excellence will liberate on how the drivers, facilitators and enablers are considered to minimized waste and deliver business value. Organizations have long recognized the importance of managing key resources such as people and information. Information has now moved to its rightful place as a key resource [1]. The emergence of transformative global economy, information system has became a necessity in businesses to obtain organizations operational excellence, adaptation to new business models, improved decision making and providing exceptional customer service, and eventual competitive advantage of the enterprise setting while keeping business alliances. Information systems are designed and "developed for different purposes, depending on the needs of human users and the business [1]. The application information system into business realized that there is a variety of disciplines employed in the "complementary of networks of hardware and software that people and organizations use to collect, filter, and process, create, distribute "[2] the data and information which are the key organizational resource. The variety of computer-based system can be deployed to organization with the help of a Systems Analyst. A System analyst is an "IT professional who specializes in analyzing, designing and implementing information systems"…he may be able to "assess the suitability of information systems in terms of their intended outcomes and liaise with end users, software vendors and programmers in order to achieve these outcomes"[3]. Information system has long been associated as tool in positioning business position in the marketplace and plays a pivotal role in the new economy.

This paper presents business sectors serviced by the pre-industry developers, explores the evolution of computer-based information system designed and developed by pre-industry system developers to, and examine reasons why there is a need for a system to countervail recurrent problems. Business owners, executives and leaders liberate their enterprise culture to change the organizations perspectives while keeping their brand promise. Companies are able to capture strengths and weaknesses of its own, interactive dashboards, real-time reports such as dashboards, and get better compared to its competitor by doing strategies rolled-out in a better, faster, cheaper, or unique when compared with rival firms.

## 2. THE CONTEXT OF THE STUDY

The author embarks this study who has been for several years' teaching the course System Analysis and Design in the university under the auspices of the College of Computer Studies and Engineering-Information Technology Department. Under this context the System Analysis and Design course, student developers dubbed as "pre-industry developers" were group to a maximum of five members. In this study, it includes eighty-five groups during the A.Y. 2011-2012 and 64 groups for A.Y. 2012-2013. The developers were tasked to scout a company to which they will study the business processes, design and develop the computer-based system. Computer-based system should be deployed over the internet using the company's domain and will cover business processes on the business operational level or higher to strategic level.

During the preliminary study, the pre-industry developers conducted their periodic visit to their respective companies to let their client elicit and elaborate their needs and problems on their existing systems. Clients of the developers' are within the schools 12km. radius proximity and/ or within Metro Manila. As depicted on the analysed documentation, the companies' existing systems are attributed to traditional file processing, office automation and transaction processing system. Each company has an established business process and a system life cycle or organizational process of developing and maintaining a system. Kendall & Kendall defined SDLC as "a phased approach to analysis and design that holds that systems are best developed through the use of a specific cycle of analyst and user activities" [1] which has been used by the developers. The figure 1 depicts the Stages of System Development Life Cycle [1].

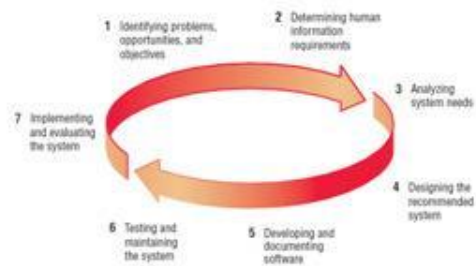

Figure 1. The seven phases of the systems development life cycle

Developers have had undergone preliminary study, feasibility study, detailed system study, system analysis, system design, coding and testing, implementation, and maintenance of the developed systems. During the system design, only the Data Flow Diagram and Data Dictionary have been instructed to use. Developers put into the handful of practice on several integrated development environment (IDE) used as platform on coding where program specifications where converted to computer instructions, and interfacing such as Java Netbeans & Eclipse, PHP & MySQL, Java Server Pages, .Net, and other technologies and frameworks for webpage design and development. The developers on S.Y. 2011-2012 developed system on the operational level which is the Transaction Processing System, while the later were tasked to developed higher level systems, such as Management Information System.

Transaction processing systems (TPS) are computerized information systems that were developed to process large amounts of data for routine business transactions such as payroll and inventory" [1] and also it " eliminates the tedium of necessary operational transactions and reduces the time once required to perform them manually, although people must still input data to computerized systems"[1]. Transaction processing is a style of computing, typically performed by large server computers, that supports interactive applications. In transaction processing, work is divided into individual, indivisible operations, called transactions…it also allows application programmers to concentrate on writing code that supports the business, by shielding application programs from the details of transaction management [6]. On the other hand, Management Information System as higher level system is defined as "computerized information systems that work because of the purposeful interaction between people and computers.. also help integrate some of the computerized information functions of a business [1], it also provides information that organizations require to manage themselves efficiently and effectively [7].

## 2.1. The Clients of Pre-industry developers

In the Preliminary portion of requirements analysis, developers were instituted to explore opportunities to working with the companies within their reach. Pre-requirements should be an organization with business licences, this is to make sure that the company is in compliance to the mandate and government regulation on business operations, and a minimum number of its employee should be fifty. To verify the veracity of the data being presented and submitted by the pre-industry developers, further documents were asked from their clients including the list of employees, organizational chart, current business process, products and services and clients of the company. Data were kept confidential for anonymity and maintaining the integrity of the developers' intention with signed confidentiality letter of agreement.

Clients of the developers were from various sectors which defines the proportion of the population engaged in the activity sector. The "primary sector of the economy extracts or harvests products from the earth. The primary sector includes the production of raw material and basic foods. [8] The secondary sector of the economy includes those economic sectors that create a finished, usable product: production and construction, secondary sector of the economy manufactures finished goods. Also, the tertiary sector of the economy is the service industry which provides services to the general population and to businesses [8]. On the positive note, quarternary and quinary sector have been served by the developers.

## 2.2. Transaction Processing System

The pre-industry developers on the context of A.Y. 2011-2012 have been directed with the provisions to develop computer-based system under Operational level of the organization. Transaction processing systems are boundary-spanning systems that permit the organization to interact with external environments [1]. Transaction processing system is an information system for business transactions which involves the collection, modification and retrieval of all

International Journal of Business Information Systems Strategies (IJBISS) Volume 3, Number 3, August 2014

transaction data, it is designed to eliminate the tedium of necessary operational transactions and reduces the time once required to perform them manually, although people must still input data to computerized systems. Transaction Processing System is attributed to as Real time systems which attempt to guarantee an appropriate response to a stimulus or request quickly enough to affect the conditions that caused the stimulus [12]. During the A.Y., pre-industry developers are tasked to design and develop a TPS which in itself means, at least they will cover a specific department of the company, identify and evaluate existing systems, identify pressing problems related to systems and current business processes, study the business process and organizational objectives and proposed system to streamline processes and countervail the problem using the information system. Initially, the process of documentation has been instituted, and the system design and development moving forward.

## 2.2 Management Information System

In this context, pre-industry developers have been challenged due to the requirements degree of complexity of their study. The same policy applies however by this time, it's more complex and by qualification and requirement itself is too big to consider. They have been invigorated to develop Management Information System which covers the organizations entire operation. Management Information System is computerized information systems that work because of the purposeful interaction between people and computers. By requiring people, software, and hardware to function in concert, management information systems support users in accomplishing a broader spectrum of organizational tasks than transaction processing systems, including decision analysis and decision making [1]. A management information system (MIS) provides information that organizations require to manage themselves efficiently and effectively [7].

Management Information System refers to the study of how individuals, groups, and organizations evaluate, design, implement, manage, and utilize systems to generate information to improve efficiency and effectiveness of decision making, including systems termed decision support systems, expert systems, and executive information systems [13]. A management information system can also help integrate some of the computerized information functions of a business [1]. Businesses uses output of Management Information System in decisions making and thereby generate new knowledge after processing information.

## 3. METHODOLOGY

The methodology used in this study is descriptive and qualitative secondary data analysis. For this study, the existing databases, reports and memoranda were utilized [4]. Most of the data have been drawn from the database of student developers of transaction processing system and management information system and IT Department Faculty Quality Circle Report were reprocessed, reclassified and analyzed. Sources of quantitative data are the group sampling and statistics, while the qualitative includes transcripts of focus group discussion, semi-structured interviews, and field notes and participant observation. Vartanian posited that analysis of secondary data, where "secondary data can include any data that are examined to answer a research question other than the question(s) for which the data were initially collected" [5].

## 4. RESULTS AND DISCUSSION
### 4.1 Sector of Economy

Based on the analyses made, there were 64 companies which were recipient during the A.Y. 2011-2012 and 85 companies during the A.Y. 2012-2013. Sectors that have been a recipient of developed Transaction Processing System includes Business Support Services (8), Call Center/BPO(1), Construction Materials(2), Defence(Government) (2), Development Finance(2), Education(22), Engineering(1), Food and Beverages(4), Healthcare(3), Industrial Manufacturing(12), Industrial Transportation(1), Insurance(1), Property(1), Retail(4). The figure 2 depicts the distribution to sectors of economy.

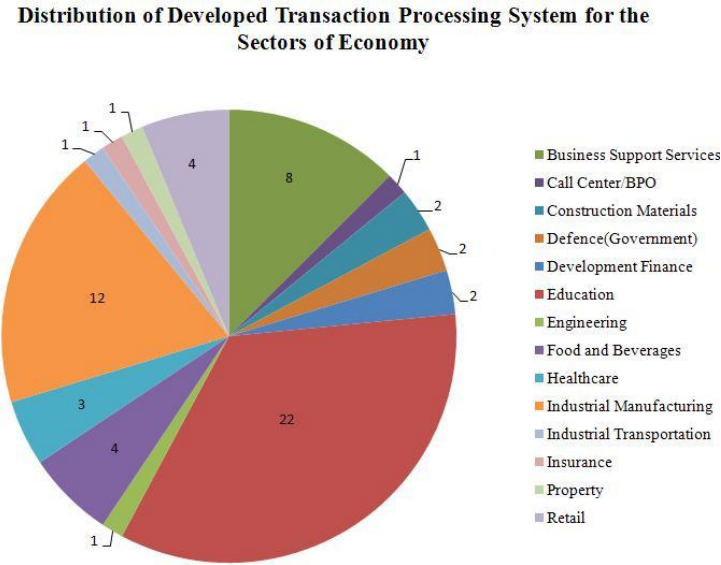

Figure 2. Distribution of Transaction Processing System per sector of economy

Based on the results, an educational institution has been the recipient with high number of required system to 22 schools herein. An educational institution is attributed to, a training institute, pre-school, elementary school, community college, and a university. In the website of the International Labour Organization says Education and training have long been recognized as key drivers of social and economic development [10] and therefore each institution with the shared vision and its mandate for its stakeholder and vow to embrace its operation and brand of promise in delivery of quality education and exceptional administrative services using computing technology. Under this study, the Transaction Processing Systems were deployed over the Local Area Network, using the company suggested system infrastructure.

While sectors which were recipients' of the studies on Management Information Systems includes; Agriculture and Agri-Processing (1), Business Support Services (6), Call Center/BPO (25), Construction Materials (2),Cooperative (3) , Education (26),Engineering (1),Financial Services (2),Food and Beverages (3), Healthcare (2), Industrial Manufacturing (6), Legal Services (1), Public Sector (2), Retail (5). From this batch of pre-industry developers, an Education sector has been the most serviced sector of economy with 26 number of educational institution. As previously noted, an educational institution is attributed to, a training institute, pre-school, elementary school, community college, and a university. A Call Center/BPO is closely second to which 25 call center and BPO companies has been serviced by the pre-industry developers. The figures 3 show sectors distribution of develop management information system. This context, the pre-industry developers has serviced at least one of sector under primary category which is under Agriculture.

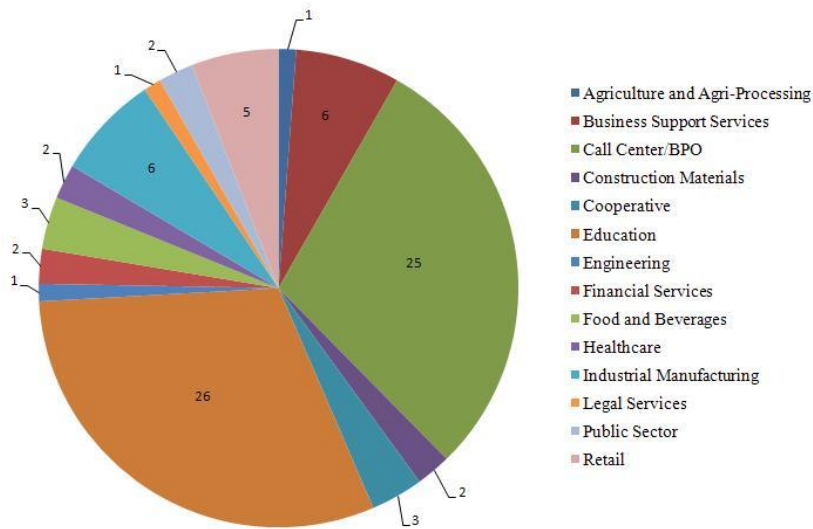

Figure 3.  Distribution of Management Information System per sector of economy

The 85 companies which were the recipients of Management Information System used their own domain during the deployment and parallel implementation. The system has been deployed over the internet and other company opted to have at least one of the module or transaction processing system implemented using Local Area Network to localized their data should access of their clients and suppliers need not to be with system functions.

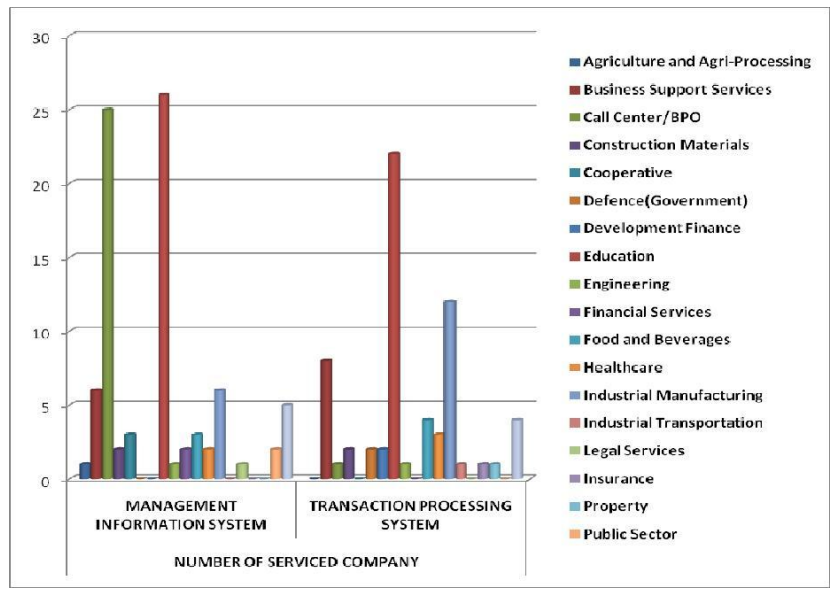

Figure 4.  Type of Sector of economy serviced by pre-industry developer

With these, each have employed different system infrastructure since some have adopted the full implementation and several institutions have had initial implementation on parallel to consider the strengths and weaknesses of the new system before fully migration. Once the new system is capable to carry out business processes, they will fully implement the developed Management Information System.

The Figure 4 presents sectors of economy which the computer-based system has been design and developed for. Based on the representation, given the two perspectives, Education sector has been serviced most which outnumbered other sectors. Furtherance of the results is depicted on the evolution of information systems.

### 4.2 Evolution of Information System

The study shows the following computer-based system that has been designed and developed by the pre-industry developers based on the needs of their clients to wit; (1) Accounting Management System, (2) Asset Management System, (3) Attendance Management System, (4) Billing Management System, (5) Clinical Management System, (6) Commissioning Management System, (7) Computing Management System, (8) Customer Relationship Management System,

(9) E-Business Management System, (10) Learning Management System, (11) Enrolment Management System, (12) Academic & Grading Management System, (13) Human Resources Management System, (14) Recruitment & Deployment Management System, (15) Order Management System, (16) Library Management System, (17) Patient Management System, (18) Payroll Management System, (19) Preventive Maintenance Management System, (20) Project Management System, and (21) Sales Management System.

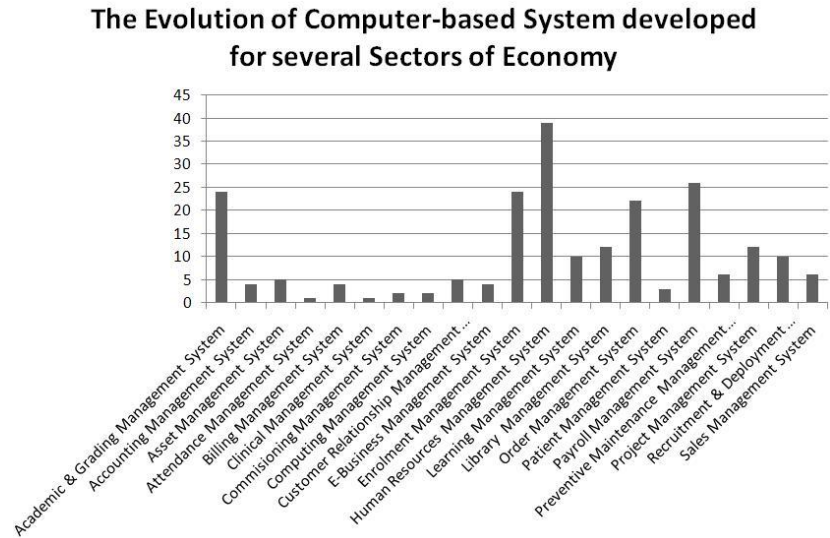

Figure 5. The Evolution of Computer-based System developed by pre-industry developers

The figure 5 shows the evolution of the systems designed and developed for their clients. Based on the result, Human Resources Management System is the most need by the companies due to its purpose of maintaining the quality of their employees' records as one of their organizational resource. In this text, subsections 4.2.1 to 4.2.10 presents the features' and common system modules that are integrated and develop to support the developers clienteles' business and organizational activities and data management initiatives. Here are the details of the Top List Computer-based Information System.

Table 1. Top List of Computer-based Information System.

| TYPE OF INFORMATION SYSTEM | # |
|---|---|
| Human Resources Management System | 39 |
| Payroll Management System | 26 |
| Enrolment Management System | 24 |
| Academic & Grading Management System | 24 |
| Order Management System | 22 |
| Library Management System | 12 |
| Project Management System | 12 |
| Learning Management System | 10 |
| Recruitment & Deployment Management System | 10 |
| Preventive Maintenance Management System | 6 |

The table 1 list down the Top 10 Computer-based information system that has been developed to numbers of sectors of economy which Human Resources Management System topped-list, as the most important information system in every organization. It's good to know that companies are keeping their HRMS as top of their priority giving an implication to managing people is the key to managing organization and business wisely. Below are the walkthrough on the system modules of some information system on the list.

### 4.2.1 Human Resources Management System

Human Resources Management System is an online business solution or software intended for human resources personnel used for data entry, data tracking, data management, analysis on information, including human resources, payroll, and management and accounting related functions within the organization. It covers management of employee information, company-employee related documents such as 201 file, benefits administration with integrated payroll, compensation and benefits of employees, employee handbook, and applicant tracking towards resume management and among others. It also provides appropriate information on attendance monitoring, paid time off policy (PTO), positions held, performance appraisal, performance evaluation, trainings, grievances and disciplinary actions, succession plans, and other when deem necessary.

The system modules have include (1) **Staffing Management**; (a) Recruitment, Applicant Tracking, Job Offer and Refusal, (2) **Equal Opportunity on Employment** including the ,(a) Action Plans, (b) Workforce Management, (c) Capability Analysis (3**) Human Resources Planning and Analysis** including; (a) Organizational Charts, (b) Projections and Staff Forecasting , (c)Staff Inventories, (d) Capacity and Capability Management, (e)Job Description, (f)Job Matching for Internal Staffing, (g) Behaviour and Absenteeism, (4) **Human Resources Development, Health, Security and Safety** ; (5) **Compensation and Benefits; (a)** Salary & Wage Administration, (b) Flexible benefits administration and management, (c) conversion and usage of institutional and government mandated leaves, (d) analysis on benefits usage(6) **Employee and Labour Relations** includes; (a) Grievances Management, (b) Records Auditing, (c) Engagement and Attitude Survey, (d) Exit Interview Analysis.

### 4.2.2 Payroll Management System

Payroll Management is defined in encyclopaedia as "the administration of the financial record of employees' salaries, wages, bonuses, net pay, and deductions. Effective payroll automation collects all relevant information in one place in electronic format; reducing mistakes by eliminating the need to synchronize and manage otherwise duplicate data sets [14]. This information system deals with the financial aspects of employee's tenure in an organization. It includes management and administration of Payroll Processes from capturing attendances of the employee to an end-to end payroll scheme. It includes generation and management of payroll and salary structures, defines emoluments, deductions, leaves and conversion of leaves to cash, incentives, generate periodic reports for internal and external submissions, and provides payslip in employees' own portal with managed security. This improves the work satisfaction of the wages clerk, as routine work is reduced [15]. Also, Payroll Management System performs the calculations and payment of salaries and wages, tax withholding settlements, submission of regulatory reports and periodic one, calculation of annual holidays and deductions management to add into the system details.

### 4.2.3 Enrolment Management System

This is used by schools and universities for a well-planned and strategic approach to eliminate periodic queues of enrolees during enrolment period; Enrolment Management System is a definite advantage as an online enrolment delivery where students and guardian can enlist their children on respective classes of a school or university system. System is decomposed to users on the Administration, Registrar, Accounting, students, faculty, enrolment team and assistant, and guidance counsellor. An enrolment process is followed however inefficient processes are eliminated. It includes , courses available for a semester or school year, subject available slots, schedules, faculty administration and distribution of teaching loads, academic records, program curriculum and courses, pre-requisites and co-requisites, enlistment to courses available, assessment and payment, enrolment reports and statistics to list down few.

### 4.2.4  Academic & Grading Management System

Some schools and universities are taking steps to liberate their culture through technology. Some of the pre-industry developers embark using information technology. Academic & Grading Management System provides portal for students, parents, registrar, faculty, deans and department chairs, guidance and shared services, accounting and among others. System decomposition includes; Course Management, Faculty Management, Client Management, Class Administration, Enrolment Administration, Capacity Allocation Management, Examination Administration, Graduation Administration, Evaluation Administration, Grant & Scholarship Management, Financial Management, Reporting Management, Security Management.

### 4.2.5 Order Management System
Order Management System is an integrated product and inventory system which encompasses modules decomposition such as; product information and catalogue, inventory, promotions and pricing, vendors management, receiving orders and goods, order entry and customer service which includes returns and refunds on allowable terms, financial processing, billing and accounts payments, order processing, including order selection, printing, picking up orders, packaging and delivery or shipping, business and data analytics and regulated reporting management, with special functions on the financial of the business including cash flows, payables, receivables, ledgering and among other intended functional areas.

Companies on distribution, sales and marketing have ask the developers to develop an Order Management System with customized modules based on their requirements, products and services being offered to the consuming public. Order Management System is used to keep track of customers, accounts, credit verification, product delivery, billing, etc. [16]. Order Management requires multiple steps in a sequential process like Capture, Validation, Fraud Check, Payment Authorization, Sourcing, Backorder management, pick, pack, ship and associated customer communications. Order management systems usually have workflow capabilities to manage this process [17].

### 4.2.6 Library Management System

Library is a collection of books, public or private; room or building where these are kept; similar collection of films, records, computer routines, etc. or place where they are kept; series of books issued in similar bindings as set [18]. The academic library provides a quiet study space for students on campus; it may also provide group study space, such as meeting rooms. In North America, Europe, and other parts of the world, academic libraries are becoming increasingly digitally oriented. The library provides a "gateway" for students and researchers to access various resources, both print/physical and digital [19].

Cognizant to the schools or university's mission of providing quality education, the educational resources are offered to some extent using a Library Management System. The purpose of having the LIS is to automate library resources and maintain the quality of work during (a) **acquisitions** such as review and ordering, receiving, and invoicing of library resources and materials (b) **cataloguing** includes classifying and indexing materials and library resources, (c) **circulation** sums up to lending materials to patrons and receiving them back) (d) **serials** (tracking magazine and newspaper holdings) the (e) use of **Online Public Access Catalogue (OPAC)** as a public interface for users or so-called patrons. The system provides ways to penalized patrons with overdue books track using the system, payments of overdue books, report management and security mechanism.

### 4.2.7 Project Management System

Project management is the process and activity of planning, organizing, motivating, and controlling resources, procedures and protocols to achieve specific goals in scientific or daily problems. A project is a temporary endeavor designed to produce a unique product, service or result [20]. The primary challenge of project management is to achieve all of the project goals [21] and objectives. In handling a project, the Project Managers and stakeholder shall consider the following primary constraints which the scope, time, quality and budget [22]. A Project Management System may be able to help the organization and project managers or researchers to capacity to help plan, organize, and manage resources and develop project estimates.

Project Management System helps to manage estimation and planning, scheduling, cost control and budget management, resource allocation and estimates, collaboration software, communication, decision-making, Quality management and documentation or administration systems [23]. Project management software is used to help remote teams keep track of essential workflow items including (a) who is assigned and responsible for a given task, (b) what is the priority of a task in relation to others, (c) comments and discussions regarding a task, and (d) description of the task [24]. Project management system is decomposed into modules such areas of project scope, project budget, project schedules, project resources, project quality, project estimates, project communication, project human resources management, project risk management, project procurement management, project stakeholders management, project milestones, task dependencies, project reports, critical path analysis, resources levelling and many others.

### 4.2.8 Learning Management System

A learning management system (LMS) is a software application for the administration, documentation, tracking, reporting and delivery of e-learning education courses or training programs [25]. A Learning Management System delivers content but also handles registering for courses, course administration, skills gap analysis, tracking, and reporting [26]. An LMS is the infrastructure that delivers and manages instructional content, identifies and assesses individual and organizational learning or training goals, tracks the progress towards meeting those goals, and collects and presents data for supervising the learning process of organization as a whole [27]. An ideal learning management system has embodied the following attributes; (a) centralize and automate administration, (b) use self-service and self-guided services, (c) assemble and deliver learning content rapidly, (d) consolidate training initiatives on a scalable web-based platform, (e) support portability and standards, (f) personalize content and enable knowledge reuse [25].

A learning management system contains modules such as; (a) Course Content Delivery, (b) Student Registration and Administration, (c) Training Event Management (i.e., scheduling, tracking), (d) Curriculum and Certification Management, (e) Skills and Competencies Management, (f) Skill Gap Analysis, (g) Individual Development Plan (IDP), (h) Reporting, (i) Training Record Management, (j) Courseware Authoring, (k)Resource Management (l)Virtual Organizations [28].

### 4.2.9 Recruitment & Deployment Management System

Businesses and other organization have been using Recruitment Management System also known as applicant tracking system (ATS) is a software application that enables the electronic handling of recruitment needs. An ATS can be implemented on an enterprise or small business level, depending on the needs of the company. An ATS is very similar to customer relationship management systems, but are designed for recruitment tracking purposes. In many cases they filter applications automatically based on given criteria such as former employers, years of experience and schools attended [29]. Applicant tracking systems may also be referred to as talent management systems (TMS).

A recruitment management system (RMS), also known as an e-recruitment or online recruitment system, is a multi-component software tool designed to automate and facilitate the processes involved in finding, attracting, assessing, interviewing and hiring new personnel [30]. Based on the developed systems, the following modules were present including but not limited to wit; the lifecycle of talent or recruitment, such as job posting, application, interview to hiring planning, recruitment process from resume management to contract and job offer, performance management, learning, career planning and development, succession planning, compensation pay grades, talent reviews and measuring and reporting management, some of the system includes engagement measures, workforce review and analyses, enterprise collaboration, vendor management and culture diversity management.

### 4.2.10 Preventive Maintenance Management System

Preventive Maintenance Management System is a computer-based system which track job orders for machine and equipment schedules and helps the workers do their jobs more effectively and to help management make informed decisions (for example, calculating the cost of machine breakdown repair versus preventive maintenance for each machine and led to better allocation of resources. It provides scheduling jobs, assigned personnel and service technicians, costs and tracks relevant information such as reported incident and problems, causes of problem, downtime issues and causes, and the mean time to resolve (MTTR). It provides interface for the inspection of a certain company asset if serviceable, operation and machine or asset breakdown. It offers scheduling, asset management, parts inventory and purchasing, including costs. It has Maintenance Plan Manager, Task List Portal, Calendar, Job Scheduler, and Graphical Scheduler.

## 4.3 Implied Significance and Importance of Computer-based System to Business

Based on the result of content analyses, there were recurrent and emerging themes derived related to advantages of the computer-based information systems to achieving relevant products and services. Business may have had used information system to business forecasting, behaviour of customers study, used business analytics, and decision making, amongst others.

### 4.3.1 Continuum process improvement

Businesses establish efforts to continuously and outgrew its competitive advantage. To seek an incremental improvement over their existence, it has charge their efficiency and effort for breakthrough to process improvement or the management of processes. With the use of computer-based information system, it has provided seamless integrated services especially on system flexibility and adaptation to the current business processes through customization of functional features and modules. In the ASQ website, the Continuous improvement process is defined as an ongoing effort to improve products, services, or processes. These efforts can seek "incremental" improvement over time or "breakthrough" improvement all at once[11] and has considered that the continuous improvement is a subset of continual improvement, with a more specific focus on linear, incremental improvement within an existing process. [11] Information system aid their business operation to streamline existing business processes, help them realized in eliminating inefficient business process and meanwhile eliminate waste in the operation and production. These wastes have been considered to be deadly on their end and costly by nature. Hence, the removal of these waste led to delivery business value where customers' view is included on the view of the system.

### 4.3.2 Business process reengineering

Information systems (IS) supports in the process of redesigning and practice of rethinking how the work is done within the business culture. Based on the reflections and interviews of the managers and information systems personnel, IS let them understand the simplified work in service delivery in relation to processes and use of information technology to fulfil its business mission to its clients. Using computer-based information system in specialized functional areas in the organization that has been automated, it has resulted at par from its previous assessment of performance on processes and productivity. The information system with customized system modules and user-centered design helped the business leverage towards optimal resources management with reduced costs. The system has been designed and has recognized sub-processes and tasks of the employees using each part of the system. Although, they have had agreed to work with the pre-industry developers just to streamline current business process, it has eventually open the door to their organization's process innovation.

### 4.3.3 Business driver and facilitator
The use of information system help the company realized that there is the necessity to assess their current situation to know where are they at the very moment that is best starting point to include analysis on their situational assessment. IS has been used to moving forward using their strategic directions. This situational assessment helps them to see the future trends in their line of business, and therefore they block new synergies to visualize and establish vision and mission statement. Moving forward, IS open an avenue for them to develop road map to fulfil their vision and mission. Everyone get to know of where they are going, and therefore the key value to strategy development will let them know how to get there.

### 4.3.4 Customer satisfaction

Measuring and improving customer satisfaction is an important function of a business. The companies are bold enough to divulge their promise to meet customers' expectation. And as a customer-focused company, they have had administered customer survey, feedbacks and other mechanism using the developed system. Surveys were administered online after transaction of each customer to rate their services in terms of quality and timeliness of their services, and quality of their products. Indicators of a satisfied customer led to loyalty to your products and services. Results showed positive outlook of customers towards companies, product, services and more projects.

## 5. CONCLUSIONS

In this $21^{st}$ century business operations, business value can be substantially delivered by using appropriate technology as tool in the delivery of products and services. Information system has served alike as reporting tools from the rank and file employees, used in knowledge management, drill-down analysis for decision making and providing customers- customers' satisfaction. Technology has provided modest aid to improved employees productivity, measure the company's current performance, study the customers' behaviour, and support system for business analytics. In Metro Manila, where business districts and highly urbanized cities are located; businesses around are still operating under the traditional file processing.

On the positive note; a university student developers can help-out by providing information systems and business software appropriate to these businesses to minimize waste in the production, encapsulate the keep their brand promise, and maintain efficient delivery of the business value. .

### ACKNOWLEDGEMENT

The author would like to thank the System Analysis and Design students especially those batches of A.Y. 2011-2012 and 2012-2013 of Jose Rizal University- Information Technology Department.

## Author

Engr. Daniel D. Dasig, Jr. is currently a member of - Universal Association of Computer and Electronics Engineers (UACEE); Member- The Society of Digital Information and Wireless Communications (SDIWC), United States of America; Member-The International Association of Engineers (IAENG); Global Member-and Philippines Chapter Member of Internet Society (ISOC). He is one of the Editorial Board-International Scientific Journal in Business Economics, Commerce and Trade 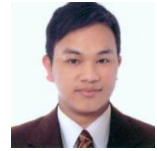 Management; Editorial Board - International Journal of Artificial Intelligence and Applications; Editorial Board- Computer Applications: International Journal; A faculty of College of Computer Studies and Engineering of Jose Rizal University, and currently holds a Technology Analyst II-Officer 3 position in TELUS International Philippines, Inc. as industry practitioner. He is a Lean Six Sigma Certified for Business Process Improvement and ITILv3 Certified in IT Service Management. He is on his thesis for Master of Science in Engineering-Computer Engineering at Polytechnic University of the Philippines-Manila, and with academic units in Master in Information System from University of Makati, Makati City.

He graduated with degree of Bachelor of Science in Computer Engineering from Samar State University, and finished a Certificate in Professional Teaching from Kester Grant College, Philippines, Inc., Quezon City. As an educator, he is actively involved in faculty development programs as member of Institute of Computer Engineers of the Philippines (ICPEP), Philippine Society of Information Technology Educators (PSITE), and Philippine Society for Study of Nature (PSSN). Author and co-authored papers presented and accepted in local and international conferences and published in referred journals. He serves as Technical Paper Reviewer of several international conferences in Busan, Korea; Kuala Lumpur, Malaysia; Ostrava, Czech Republic; Bangkok, Thailand; Chennai India; and Islamabad, Pakistan.